\documentclass[aps,prl,showpacs]{revtex4}
\usepackage{graphicx}

\begin{document}

\title{Dislocation induced ac Josephson effect in high-$T_c$
superconductors}
\author{Sergei Sergeenkov and Marcel Ausloos}
\affiliation{SUPRAS, Institute of Physics, University of Liege,
B-4000, Liege, Belgium}

%\date{\today}

\begin{abstract}
A possible scenario for an ac Josephson effect initiated by the
 flow of dislocations through a mechanically loaded but electrically
unbiased superconductor is proposed. The characteristic voltages
due to the
motion of dislocations in loaded (under the applied stress of
$10^7N/m^2$)
$YBCO$ crystals are estimated to be of the order of a few picovolts
which corresponds to the Josephson frequency of $10 kHz$.
\end{abstract}


\maketitle

According to recent experimental observations [1-5], high-$T_c$
superconductors (HTS) possess a large number
of extended defects which give rise to the various interesting
phenomena in
these materials [6-11]. In particular, considering a twinning
boundary (TB) as
insulating region of the superconductor-insulator-superconductor (SIS)
structure [6], the possibility of
dislocation-induced dc Josephson effect in HTS has been recently
discussed [10,11].
At the same time, using the so-called method of acoustic emission
(see, e.g.,
Ref.12), a quite tangible motion ( flow) of twinning dislocations in
HTS crystals (at $T<T_c$) under external load has been registered
[13-15].

Based on these experimental findings, in the present paper a
possible scenario
for an ac Josephson effect related to the external load induced  flow
of dislocations through an electrically unbiased superconducting
sample is proposed and its
realization in HTS crystals is discussed.

As is well-known [16],  a constant voltage $V$  applied to a
Josephson
junction causes a time evolution of the phase difference between
two superconductors, $d\theta /dt=2eV/\hbar$. As a result an ac\
Josephson current occurs through such a contact
\begin{equation}
I_s^V(t)=I_c\sin (\theta _0+\omega _Vt),
\end{equation}
where $\theta _0$ is the initial (at $t=0$) phase difference,
$\omega _V=2eV/\hbar$ the Josephson frequency, and $I_c$  the
critical current.

In the present paper we discuss another
possibility for ac Josephson effect which is due to the external
load ($\sigma$) induced  flow of dislocations through an unbiased
($V=0$)
superconducting sample. Namely, we assume [10,11] that initially
(at $t=0$) there is a twin boundary (which is characterized by a
non-zero dislocation-induced  deformation $\epsilon $ [8]) inside a
superconductor which creates a SIS type single junction Josephson
contact.  When a  mechanical stress is applied to the system,
causing a  flow of dislocations (and thus the corresponding
displacement of the insulating layer which is created by these
dislocations, see Fig.1) through a loaded crystal, a time-dependent
phase difference $d\theta /dt =(d\theta /d\epsilon )
\stackrel{.}{\epsilon }$ [where $\stackrel{.}{\epsilon }=d\epsilon
/dt$ is the rate of plastic deformation under an applied stress]  is
expected to occur in such a {\it moving} contact. For simplicity, in the
present paper we postulate a linear dependence for the induced
phase difference assuming $\theta  (\epsilon )=A\epsilon $ (where
$A\simeq 1$ is a geometrical factor).
To stay within a short junction approximation (for which Eq.(1) is
valid), we assume also a constant  (time-independent) rate of flow
of dislocations through a loaded crystal. Finally, taking into account
the dependence of $\stackrel{.}{\epsilon  }$ on the number of
dislocations (of density  $\rho $) and a mean dislocation rate $v_d$,
viz. [17] $\stackrel{.}{\epsilon }$=$b\rho v_d$ (here $b$ is the
value of the Burgers vector), the dislocation-induced zero-voltage
ac Josephson current reads
\begin{equation}
I_s^{\sigma }(t)=I_c\sin (\theta _0 +\omega _{\sigma }t)
\end{equation}
Here $\omega _{\sigma }=b\rho v_d(\sigma )=2eV_d(\sigma
)/\hbar$, where $V_d(\sigma )=\hbar b\rho v_d(\sigma )/2e$ is a
characteristic voltage
due to the motion of dislocations, and
$\sigma$ is the external stress which causes the
 flow of twinning dislocations.
Thus, comparing Eqs. (1) and (2), we conclude that one can observe
a dislocation-induced ac Josephson effect either by applying a
constant voltage to the (immobile) contact or, alternatively, by
applying a mechanical stress to electrically unbiased but mobile
[due to  flow of dislocations (forming the insulating layer of SIS
type contact) through the loaded crystal] Josephson junction.
Regarding the latter possibility, it is interesting to mention that
according to the Faraday's law of induction, a voltage induced in a
closed circuit can be presented as a rate of magnetic flux flow
through this circuit, namely $V_{ind}\propto d\Phi /dt$, implying a
linear flux dependence of the phase difference through the contact,
i.e. $\theta  (\Phi )=2e\Phi /\hbar$ .

\begin{figure}[ht]
\begin{center}
\includegraphics[width=12cm]{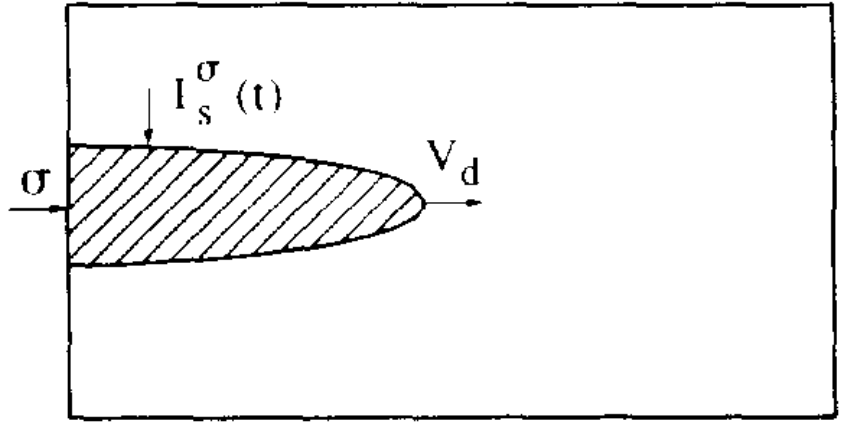}
\end{center} \vspace*{-10mm}
\caption{Sketch (out of scale) of the dislocation induced ac
Josephson effect geometry. The  flow of twinning dislocations of
density $\rho $ (with $v_d(\sigma )$ being the rate of the growing
twin boundary tip) under the external stress $\sigma$ results in
ac Josephson current $I_{s}^{\sigma }(t)=I_c\sin (\theta _0
+\omega _{\sigma }t)$, with the characteristic frequency $\omega
_{\sigma }=b\rho v_d(\sigma )$ (the insulating region due to
moving dislocations is hatched).}
 \end{figure}

Since [17] $v_d\sim (\sigma /\sigma _m)^n$ with $n\simeq
1$ and $\sigma _m$ being the so-called ultimate stress, according
to Eq.(2) the ac Josephson effect due to  flow of dislocations
disappears when the external load is relieved ($\sigma \rightarrow
0$).  At the same time, the dc Josephson effect can still exist
provided there are enough dislocations in a superconductor to
sustain the insulating (or normal metal) region for SIS (or SNS) type
contacts [an explicit dislocation density dependence of the initial
phase difference $\theta _0$ and the critical current $I_c$ of a dc
effect has been discussed in Refs.10 and 11, respectively].

Let us estimate the order of magnitude of the dislocation-induced
voltage $V_d$. To this
end, we have to know the order of magnitude of the dislocation rate
$v_d(\sigma )$. According to recent investigations [14,15],
based on the method of acoustic emission [12], the  flow of twinning
dislocations with the maximum rate of $v_d=0.1m/s$ has been
registered in
$YBCO$ crystals at $T=77K$ and under the external load of $\sigma
=10^7N/m^2$
[with the ultimate stress of $\sigma _m\cong 10^8N/m^2$]. Taking
$\rho _m\cong 10^{14}m^{-2}$ and $b\cong 1nm$ for the
maximum density of dislocations, and the magnitude of the Burgers
vector in heavily dislocated
$YBCO$ crystals [5,8], we get $V_d(\sigma =10^7N/m^2)=\hbar
b\rho v_d/2e\cong 1pV$ for the order of magnitude of the
characteristic voltage
due to the motion of dislocations, which is equivalent to the
characteristic Josephson frequency $\omega _{\sigma }\cong 10
kHz$. It would be quite interesting to observe the above-
discussed effect in dislocated HTS crystals experimentally.

In summary, a possible scenario for an ac Josephson effect
originating from
the external load induced  flow of dislocations through an
electrically unbiased
superconducting sample has been proposed. The characteristic
Josephson frequency due to the
motion of dislocations in loaded (under the applied stress of
$10^7N/m^2$)
$YBCO$ crystals is estimated to be of the order of 10 kHz.

\end{document}